\newcommand{\4}{$^4$He}
\newcommand{\3}{$^3$He}
\newcommand{\tc}{T$_C$}
\newcommand{\tl}{T$_\lambda$}
\newcommand{\etal}{{\it et al.}}
\documentstyle[aps,twocolumn]{revtex}
\begin{document}
\renewcommand{\textfraction}{0.5}
\renewcommand{\topfraction}{0.5}
\renewcommand{\bottomfraction}{0.0}
\setcounter{topnumber}{1}
\setcounter{bottomnumber}{0}

\wideabs{
\title{Observation of the Lambda Point in the \4-Vycor System: A Test of Hyperuniversality}
\author{G.M. Zassenhaus and J.D. Reppy}
\address{Laboratory of Atomic and Solid State Physics and the Cornell Center for Materials Science, 
Cornell University, Ithaca, New York 14853}

\date{\today}
\maketitle

\begin{abstract}
We have performed a high resolution specific heat measurement on \4 completely filling the pores of Vycor glass. Within 10mK of the superfluid transition we have found a peak in the heat capacity which is only 0.02\% the size of the background. The peak can be fit with a rounded version of the ``logarithmic singularity'' observed in bulk \4. Along with the previously observed ``2/3'' power law dependence of the superfluid density on temperature, this strongly suggests that the disorder imposed by the Vycor is irrelevant to the 3DXY superfluid phase transition. The critical amplitude of the peak, while in agreement with that found in other experiments in dilute superfluid \4, is considerably larger than that predicted by the theory of hyperuniversality.
\end{abstract}

}

One of the foundations of the modern theory of critical phenomena is that all phase transitions are divided into a handful of universality classes, grouped together on the basis of a common symmetry and dimensionality in the ordered phase. Experimentally, one groups phase transitions on the basis of the powerlaws in the reduced temperature $t = 1 - T/T_C$ which thermodynamic properties follow very near to the transition temperature \tc. The critical exponents of these powerlaws, which can be predicted to high accuracy thanks to the computational machinery of the renormalization group theory (RGT), are the same for all systems within a universality class. 

Because of its high purity, lack of strain, and the relative ease of precision measurement, the superfluid transition of \4~is an unrivalled testing ground not only for determining the values of critical exponents but also for answering questions about universality. Very precise measurements of the specific heat of \4~to within a few nK of \tl, the critical temperature of the superfluid transition, have confirmed the powerlaw 

\begin{equation}
C = \left\{ \begin{array}{ll} 
A(-t)^{-\alpha} + B & {\rm for} \hspace{0.12in} T>T_\lambda \\
A^\prime t^{-\alpha} + B^\prime & {\rm for} \hspace{0.12in} T<T_\lambda ,\\
\end{array}
\right.
\label{eqn:cvpower}
\end{equation}

where $\alpha = -0.01245$, $A=5.594$ J/mol K, and $A/A^\prime = 1.054$. $B$ and $B^\prime$ are non-universal parameters.\cite{Lipa,Singsaas} This behavior is observed over 8 orders of magnitude in reduced temperature.\cite{Lipa} 

The superfluid density has also been measured precisely and under the conditions of saturated vapor pressure is found to obey the powerlaw 

\begin{equation} 
\rho _{S} = \rho _{S0}t^{-\nu} \hspace{0.12in} {\rm for} \hspace{0.12in} t<0,
\label{eqn:rhopower}
\end{equation}

where the critical exponent $\nu=0.6702$ and the critical amplitude $\rho_{S0}=0.351$ g/cm$^3$.\cite{Goldner} In both experimental quantities, the agreement between the experimentally determined critical exponents and precise RGT calculations is a stunning confirmation of our understanding of critical phenomena in pure systems.\cite{Zinn-Justin} The ``logarithmic singularity'' in the specific heat of equation \ref{eqn:cvpower} and the ``2/3 power law'' of equation \ref{eqn:rhopower} identify the superfluid transition of \4 as a member of the 3DXY universality class. 

Another pillar in the theory of critical phenomena is the theory of hyperuniversality.\cite{Hohenberg} Hyperuniversality is a statement that the amount of energy (measured in units of $k_B$T) per fluctuation is the same for all systems within a universality class. This is expressed by defining the universal constant $R$

\begin{equation}
R = \left(\frac{A}{k_B}\right)^{\frac{1}{3}}\xi_0,
\label{eqn:hyperuniversality}
\end{equation}

where $\xi_0$ is the critical amplitude of the correlation length. RGT calculations find $R=0.96$.\cite{Hohenberg} While $\xi$ is not accessible directly, it can be calculated from the superfluid density by using the Josephson relation\cite{Fisher}

\begin{equation}
\xi_0 = \frac{k_B T_C m_{He}^2}{\hbar ^2 \rho_{S0}}.
\label{eqn:josephson}
\end{equation}

The relation in equation \ref{eqn:hyperuniversality} was tested by Ahlers and co-workers, who performed specific heat and superfluid density measurements near $T_\lambda$ in samples held at a series of pressures between saturated vapor pressure (SVP) and 30 bar.\cite{Singsaas}. Because the order parameter is expected to have the same symmetry regardless of pressure, \4 samples along this $\lambda$-line are believed to all be in the 3DXY universality class. Ahlers and co-workers were able to adjust $\xi_0$ from 3.1$\AA$ to 3.5$\AA$ by varying the pressure, and in this range they found a pressure-independent value of $R=0.84$, in reasonable agreement with the theory of hyperuniversality.   

Since the range of pressures over which superfluid \4~exists is restricted from 0.05 bar (SVP) to 32 bar, one can only adjust the correlation length by 13\%. On the other hand, by imposing disorder on the \4, it is possible to test the theory with samples where the $\xi_0$ 

\begin{figure}[t!]
\parbox{2.6in}{
\includegraphics{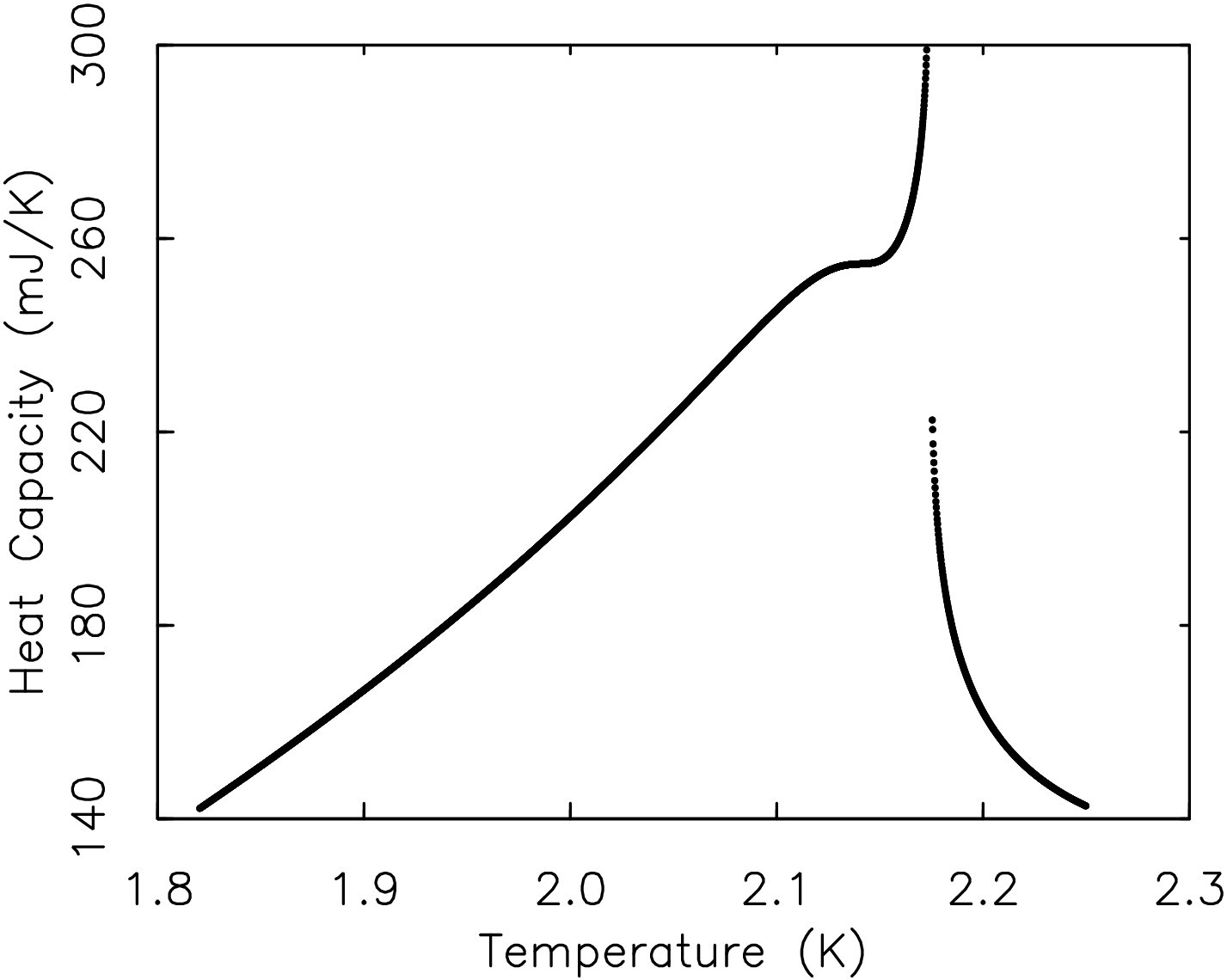}
\vspace{2.4in}
\label{fig:totalcv}
}
\caption{The total heat capacity of the cell.}
\end{figure}

exceeds 100$\AA$. We can realize disorder in superfluid \4~by adsorbing it into a porous medium. According to the Harris criterion\cite{Harris}, provided that the disorder in the porous medium is not correlated on length scales larger than the correlation length $\xi$(t), the disorder is irrelevant and the disordered system remains in the same universality class as the pure system.
 Additionally, the Harris criterion demands that in the pure system $\alpha<0$, which is the case for superfluid \4.

Several porous media have been introduced into \4 so as to impose disorder. For instance, Vycor is a porous glass with 30\% of its volume being open and an average pore size of 70$\AA$. Scattering measurements show that Vycor is uncorrelated on all length scales above 100$\AA$.\cite{Vycor} Porous gold is very similar in structure to Vycor, however the porosity is about 70\% and the average pore size is on the order of 1000 $\AA$.\cite{Yoon} Both Vycor and porous gold are examples of disorder exhibiting the ``2/3'' power law in the superfluid density.\cite{Yoon,Bishop} Furthermore, cusps in the specific heat coincident with superflow onset were seen for {\em films} of \4~on Vycor\cite{Finotello} and for porous gold filled with \4.\cite{Yoon} These two facts are consistent with these disordered systems also being members of the 3DXY universality class. The picture which emerges from these studies, consistent with the Harris criterion, is that the primary effect of these porous media is to dilute the superfluid without changing the nature of the superfluid phase transition. 

What was missing from this picture, until recently, was a peak in the specific heat at \tc~in filled-pore Vycor. The failure of several attempts\cite{Finotello,Joseph,Brewer} at seeing this feature is often explained through the theory of hyperuniversality, which argues that the cusp would have been too small to see. The lack of such a peak would be inconsistent with our regarding the \4-Vycor system as another member of the 3DXY universality class. In this paper we report on the observation of the cusp near the $\lambda$-point in the \4-Vycor system, demonstrate that this system is a dilute 3D superfluid in the 3DXY universality class, and discuss its implications to the theory of hyperuniversality. 


To this end we have performed a high resolution measurement of the specific heat of the \4-Vycor system. Our best measurements have a resolution of about 5ppm. By comparison, the measurements of Finotello {\it et al.} have a resolution of about 0.2\%. The precision of our measurements exceeds that of prior measurements by a factor of 200. This comes with the additional benefit of a higher density of data points per temperature interval. While we give a brief summary of our novel technique here, a more thorough discussion of our calorimeter appears in another article.\cite{Zassenhaus} In short, our measurement takes the high resolution thermometers of Lipa and co-workers\cite{Lipa2} as a point of departure. These authors used the paramagnetic salt Cu(NH$_4$)$_2$Br$_4$ (CAB) as a thermometric element. This salt has a Curie temperature of about 1.8K, so that the magnetic flux contained in a pick-up coil around the CAB changes rapidly as a function of temperature around 2K. The SQUID-based detection system of Lipa and co-workers, through measuring this magnetic flux, allowed a temperature resolution of nearly 0.1nK at \tl. Our modification was the insertion of a resistor into the superconducting flux loop of the HRT containing the SQUID input coils, the leads, and the CAB pick-up coils. As a result, our detection system is sensitive not to the magnetic flux in the pickup coils but rather to the changes in the flux with respect to time. This is directly proportional to the temperature drift rate, \.{T}. Our system has a resolution of about 15pK/S in \.{T}.

We perform a drift measurement of the heat capacity. A constant heat input \.{Q} is applied to the cell, producing continuous warming over the temperature range from 1.8 to 2.2K, which spans the known \tc~for the superfluid transition in this sample. The heat is generated by Joule heating a low temperature resistor on the cell with a high-precision current source. For these measurements our heating rates varied from $0.2\mu$W to $5\mu$W which corresponds to temperature drift rates between 1.5$\mu$K/s and $40\mu$K/s at 2K. We calculate the heat capacity from

\begin{equation}
C = \frac{\dot{Q}}{\dot{T}}.
\end{equation}

The internal time constant of the cell was measured at various temperatures and found to be around 40 seconds. Rounding of features in the heat capacity due to the drift ranges from 0.06 to 1.6mK. 

A cold valve on the cell allows it to be filled and closed off without leaving \4 in the fill capillary. Two regulated, weakly-coupled isothermal stages absorbed heat leaks before they could make their way to the cell, to which heat flows from the final regulation stage with a thermal time constant of about 10$^5$ seconds. We measured the temperature of the experiment with a carbon glass thermometer for ${\rm T}>1.4$K and with a germanium thermometer for ${\rm T}<1.4$K. Both thermometers were checked using \tl~of the bulk helium in our cell as a reference. Our experiment was mounted on a \3 refrigerator.

In these experiments we measured the heat capacity under saturated vapor pressure of \4 filling the pores of the Vycor. The two Vycor samples are disks of diameter 2.5 cm and thickness 1 mm. We ground our specimens from a plane equidistant from the surface and the leach plane in the center of 0.953 cm thick plates. Our hope was that the samples, while still fairly large, would have fast thermal equilibrium times and fairly uniform pore structures. The samples have a volume of 0.478 cm$^3$. 

The total heat capacity of the filled cell near 2K is shown in figure 1. We can see clearly the broad peak in the heat capacity at 2.1K, well above $T_{C}$, which was first observed by Champeneay and Brewer.\cite{Brewer} Also clearly shown is the sharp cusp at the bulk $T_\lambda$. The \4 in the Vycor pores accounts for 35\% of the heat capacity of the cell at 2K. The superfluid bulk \4 is about 45\%, and the remainder is the heat capacity of the cell construction materials, including copper, brass, niobium, and the CAB salt. From carefully filling the Vycor during superfluid density measurements, we conclude that 5.33 x $10^{-3}$ moles of \4 fill the Vycor. 

\begin{figure}[t]
\parbox{2.8in}{
\includegraphics{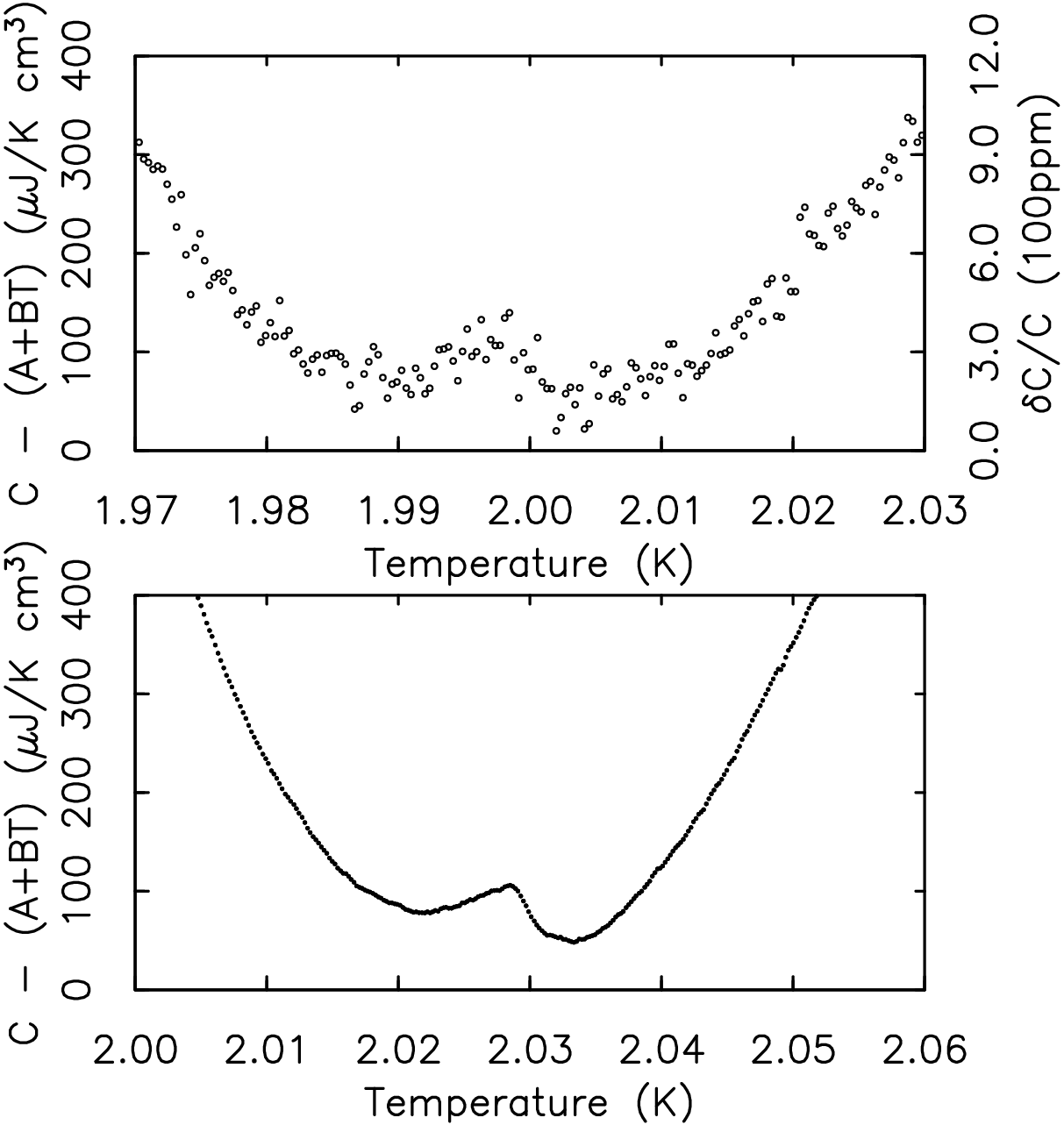}
\vspace{2.6in}
\caption{A blowup of the region of interest in the heat capacity. In both panels, a straight line has been subtracted from the total cell heat capacity so as to emphasize the cusp occuring near \tc. The upper panel shows the heat capacity from a sample with \tc = 1.995, while the lower shows a different sample with a \tc=2.031. The data in the lower panel were acquired with a more precise technique.}
\label{fig:closeups}}
\end{figure}

It is very difficult to make out a tiny feature at 2K from looking at figure 1. In figure \ref{fig:closeups} we subtract a straight line from total cell heat capacity in the region of interest for both samples. The top panel shows an early measurement on our first sample of Vycor. Using the Helmholz resonator technique\cite{Helmholz}, we measured the onset of superflow in a 1mm diameter by 3 mm long cylinder of Vycor taken from the same plate as the first sample. In this experiment a different set of thermometers were used but we nonetheless determined superflow onset to be at 1.987K, while the cusp in the specific heat is near 1.995K. The lower panel shows the most recent measurement of the second sample which was ground from a different plate of Vycor. This measurement was undertaken with a much improved calorimeter, as evidenced by the dramatic reduction in noise-to-signal ratio. After completion of the heat capacity measurements, this very same sample was loaded into a torsional oscillator cell along with the same thermometers. We found superflow onset to occur at 2.023K. From this figure we can conclude that both peaks occur within 10mK of the superflow onset temperatures as measured in the separate experiments. 

These data are of sufficient quality to allow comparison to the bulk \4 $\lambda$-peak and thereby test the theory of hyperuniversality. As with Finotello {\it et al.}\cite{Finotello}, background subtraction in this situation is very difficult. We expect the peak to have a nearly logarithmic shape to it, which means that the singular contribution we wish to keep contributes significantly (on our scale) to the heat capacity even 50mK or so away from \tc. So we took a different approach and decided instead to attempt to subtract a rounded and reduced version of the specific heat from equation \ref{eqn:cvpower}. We smoothed the heat capacity by convolving the bulk form with a normalized Gaussian function of width $\sigma$. We reduced the amplitude of the specific heat cusp by a factor $\Lambda$. Finally, we fixed the non-universal constants $B=B^{\prime}=0.4335$ J/K cm$^3$, the same values found in bulk.\cite{Ahlers}.

\begin{figure}[t]
\parbox{2.6in}{
\includegraphics{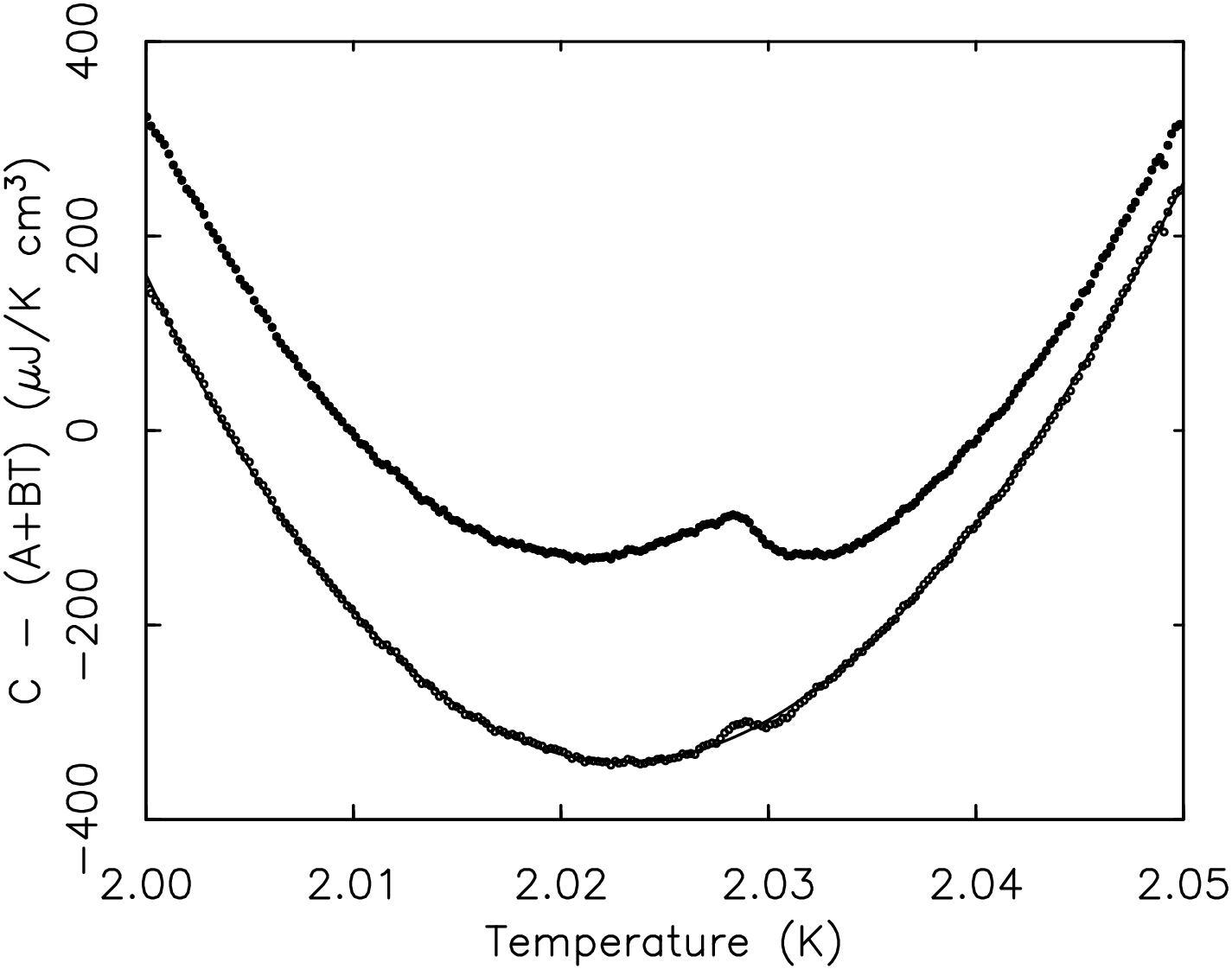}
\vspace{2.4in}
\caption{The region of the heat capacity very close to \tc with a linear fit subtracted. The top curve shows the data prior to subtracting the rounded cusp, the lower afterwards.} 
\label{fig:background}}
\end{figure}
 
We went about performing the fit by varying $\sigma$, $\Lambda$, and $T_C$ and subtracting the rounded bulk cusp from the raw heat capacity, leaving us with a residual curve. For the goodness-of-fit parameter we used the deviation from a third-order polynomial fit to the residual. Our best fit is shown in figure \ref{fig:background}. We find $T_C=2.0318\pm0.0007$K, $\sigma=2.3\pm0.8$mK, and $\Lambda$=$1.4 \pm 0.2{\rm x}10^{-4}$. The 2.3mK rounding is also very reasonable given the rounding in the superfluid density also observed in this sample. 

We see from figure \ref{fig:background} that there is in fact a large difference between the data trace and the residual even far away from \tc. Furthermore, we see that most of the bump appears to be gone, but there are still some wiggles slightly above \tc. Subtracting the best-fit third-order

\begin{figure}[t]
\parbox{2.5in}{
\includegraphics{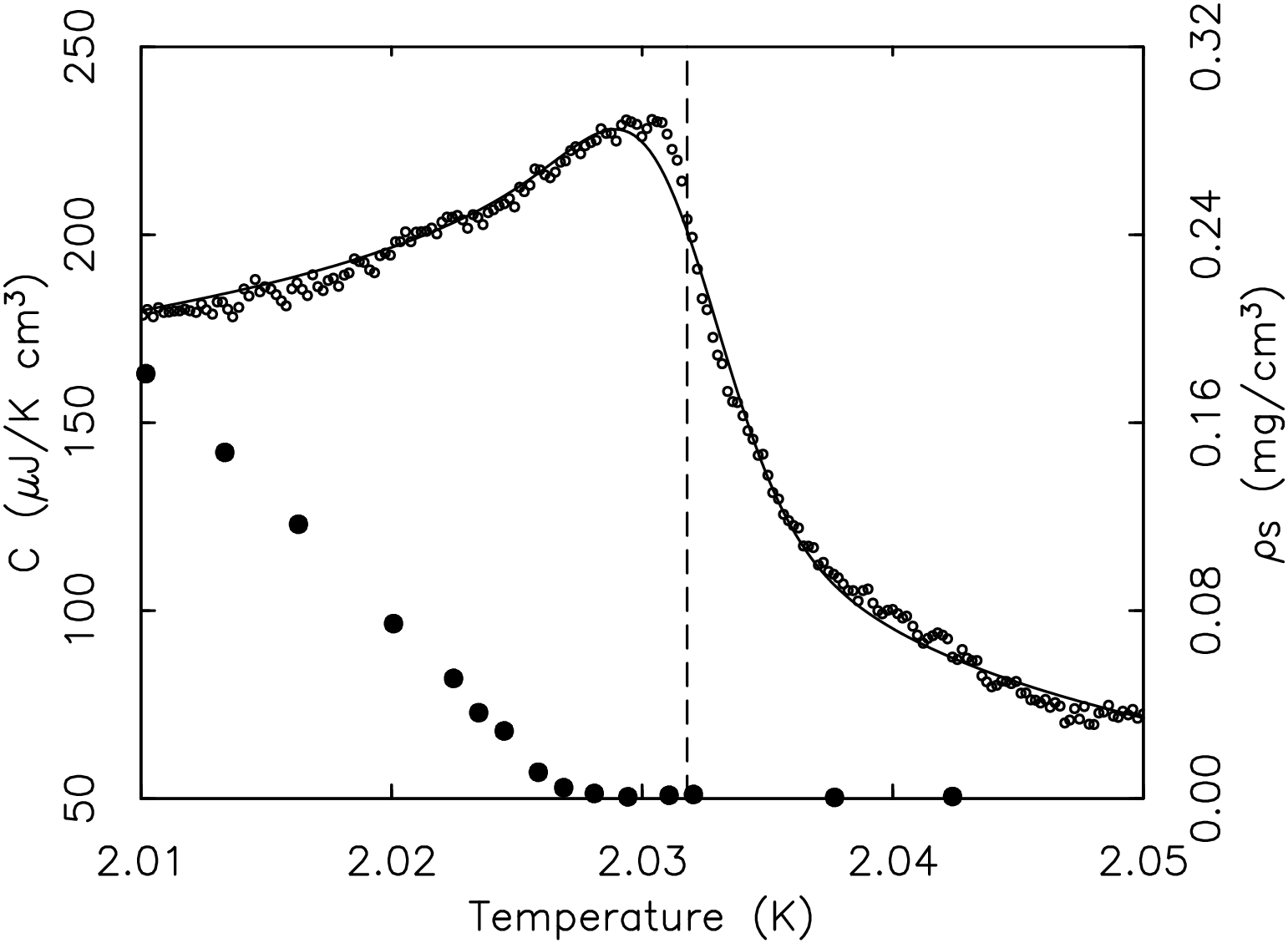}
\vspace{2.3in}
\caption{The data near \tc and the rounded cusp fit plotted with the superfluid density.}
\label{fig:cusp}}
\end{figure}

polynomial as a background function from the data trace leaves us the singular part of the specific heat. These data are plotted out to a range of 0.015 in reduced temperature in figure \ref{fig:cusp}. This is the approximate range over which the ``2/3'' powerlaw fits the superfluid density in Vycor. We see that except for the slight deviations just above \tc, the fit is fairly reasonable. We also plot the superfluid density measured in this sample and see that it goes to zero very close to the \tc~obtained from the fit.

Figure \ref{fig:cusp} lends the most striking support to the assertion that the \4/Vycor system is in the 3DXY universality class along with bulk \4. We use the predictions of hyperuniversality\cite{Hohenberg} to relate the specific heat and the correlation length. We have measured the superfluid density $\rho_S$ in this sample and found that the critical amplitude $\rho_{S0}=11.9\pm 0.1$ mg/cm$^3$, which corresponds to $\xi _0=93 \AA$. Inserting our values for $A$ and $\xi _0$ into equation \ref{eqn:hyperuniversality} we find that $R=1.2\pm0.2$. This is considerably larger than that found from measuring the specific heat and superfluid density in bulk \4 samples along the $\lambda$-line.\cite{Singsaas} However, this discrepancy is supported by applications of hyperuniversaliy to thin films of \4 on Vycor \cite{Finotello} and also to \4 filling the pores of porous gold.\cite{Yoon} We must note that particularly in the Vycor systems, we are expecting hyperuniversality to work over a factor of 30 in correlation length. One might conclude from this that the theory captures much of the physics, but either there is some fine-tuning necessary, or there is some subtle aspect of the imposition of disorder which has been ignored.

In this Paper we have shown the results of a calorimetric measurement of the \4-Vycor system over two orders of magnitude more precise than the prior measurements. The primary result of this work is the resolution of a heat capacity cusp very near to superflow onset. We have fit this cusp to a rounded and diminished cusp of the sort found at $T_\lambda$ in bulk \4. This peak allows us to test the theory of hyperuniversality and we find that our peak, although larger than that predicted by the theory, is in fair agreement with peaks measured in other disordered \4 systems. We hope to see a resolution to the remaining discrepancy between the theory and our observations. 

\acknowledgments

We wish to thank A. Tyler and  A. Woodcraft for contributing to the development of the calorimeter. We thank A.C. Corwin and J. He for their assistance in performing these measurements. We have benefitted from conversations with M.H.W. Chan, F.M. Gasparini and T.C.P. Chui. This work was carried out under NSF grant DMR96-23694 and with funding from the Cornell Center for Materials Research under NSF grant DMR96-32275.

\end{document}